# The new face of the activation volume in glass forming systems

Aleksandra Drozd-Rzoska

Institute of High Pressure Physics Polish Academy of Sciences,

ul. Sokołowska 29/37.01-142 Warsaw, Poland

**Abstract:**

This report proposes the solution of the long standing problem regarding the key property governing the pressure related dynamics in glass forming and complex systems: the determination of the apparent activation volume $V_a$ and its pressure dependence. Results are tested in the previtreous domain, up to $P \sim 1.2$ GPa, for glass forming diisobutyl phthalate (DIIP, low molecular weight liquid), isoctyloxycyanobiphenyl (8*OCB, liquid crystal), bisphenol A/epichlorohydrin (EPON 828, epoxy resin).

**Key words:** glass transition, high pressures, dynamics, activation volume,

The glass transition is the unique phenomenon associated with far previtreous 'anomalies' of dynamic properties, showing similar scaling patterns in such different systems as low-molecular weight liquids, resins, polymers, ...., liquid crystals or plastic crystals [1-3]. The key hallmark of these extraordinary phenomena are the super-Arrhenius (SA) slowing down of the structural relaxation time ($\tau(T)$, viscosity $\eta(T)$, electric conductivity ($\sigma(T)$) or diffusion $d(T)$ on cooling towards the glass temperature ($T_g$) and the analogous behavior for the isothermal compression up to the glass pressure ($P_g$) [3]. These anomalies are coupled to temperature changes of the apparent activation energy $E_a(T)$ and pressure related variations of the apparent activation volume $V_a(T)$, respectively [3-5]. Focusing on the structural relaxation time (primary, alpha), such picture is associated with the super-Arrhenius (SA) behavior of mentioned dynamic properties. Focusing on the structural relaxation time [1-5]:

$$\tau = \tau_0 \, exp\left(\frac{E_a(T)}{RT}\right) \tag{1}$$

where $P = const$ (most often: $P = 0.1 MPa$), $T > T_g$, $R$ is the gas constant and $\tau_0$ is the prefactor; the glass temperature is associated with the condition $\tau(T_g) = 100 s$.

For $E_a(T) = E_a = const$ one obtains the basic Arrhenius equation, enabling the estimation of the activation energy via $E_a' = E_a/RT = d\,ln\,\tau(T)/d(1/T)$ [2, 3]. Such calculation is not possible for the apparent activation energy: $E_a'(T) \neq d\,ln\,\tau(T)/d(1/T) = H_a'(T) = H_a(T)/RT$ [1, 6], where $H_a(T)$ denotes the apparent activation enthalpy. As shown in ref. [7], one can reach event values $H_a(T)/E_a(T) > 10$ for $T \to T_g$ in very fragile glass formers. Notwithstanding, eq. (1) seems to offer a simple way for the 'direct' estimation of the apparent activation energy: $E_a(T) = RT\,ln(\tau(T)/\tau_0)$. The application of such route requires the a priori knowledge of the prefactor: 'the universal' value $\tau_0 = 10^{-14} s$ is assumed [8]. However, estimated in this way values of $E_a(T)$ are notably biased because $\tau_0$ can range between $10^{-10}\,s$ and $10^{-16}\,s$ in different glass formers [1, 6]. The non-biased way of $E_a(T)$ determining was proposed in ref. [1], although the route is associated with notable analytical challenges. The SA eq. (1) cannot be directly applied for portraying experimental data due to the unknown general form of $E_a(T)$ and then 'substitute relations' for $\tau(T)$ or $\eta(T)$ parameterization are used [1-4]. When considering the behavior of dynamic properties on compressing, the pressure – related counterpart of the SA relation is considered [4, 6]:

$$\tau = \tau_0^P \, exp\left(\frac{V_a(P)\times P}{RT}\right) \qquad (2)$$

where $T = const$ and $P < P_g$; the glass pressure is associated with the condition $\tau(P_g) = 100\,s$. This relation simplifies to the form of the equation originally proposed by Barus [9] for viscosity, $\eta(P) = \eta_0\, exp(\alpha P)$, assuming $V_a(P) = V_a = const$: consequently eq. (2) can be recalled as the super-Barus (SB) relation. As mentioned above the determination of $E_a(T)$ is a puzzling task [1], but the situation for the pressure-related case seems to be comfortable: the estimation based on the relation $V_\#(P) = RT[d\,ln\,\tau(T)/dP]_T$, is commonly accepted [5, 10-26].

This report shows that $V_\#(P) \neq V_a(P)$, and $V_\#(P)$ is 'only' the measure of the apparent activation fragility. The new and reliable way of $V_a(P)$ determining is proposed. It is determined for supercooled, vitrifying, diisobutyl phthalate, epoxy resin EPON 828 and liquid crystal 8*OCB. Obtained $V_a(P)$ experimental data are fairly portrayed by the relation taking into account impacts of the stability limit under negative pressure and the singular pressure $P^* > P_g$.

Studies in which the apparent activation volume $V_\#(P)$ in the previtreous domain is discussed [5, 13-26] most often recalls the analysis reported by Williams [10] 1964 or Whalley (1964) [11], which can be rewritten as $V_a = d\ln\tau/dP$. Regarding, its origins one also recall the earlier report by Whalley (1958, [12]) on the impact of pressure on reaction rates. However, these reports recall the Eyring equation as the base and neglect pressure dependences of derived parameters in considered ranges of pressures [12]. Discussing, the pressure dependent apparent activation volume $\Delta V_{\#}(P)$ in glass forming liquids worth focusing is the SB eq. (2), which leads to the relation:

$$RT\frac{d\ln\tau(T)}{dP} = V_a(P) + P\frac{dV_a(P)}{dP} \qquad (3)$$

The coincidence between $V_{\#}(P)$ and $V_a(P)$ occurs only assuming $dV_a/dP \to 0$ or $P \to 0$. This is possible only low pressure, where the basic Barus equation [9] ($V_a = const$) may be valid within the limits of the experimental error. It is noteworthy that the indication $V_{\#}(P) \neq V_a(P)$ agrees with the fundamental analysis by Beyeler and Lazarus (1971, [27]), focused on diffusion processes during chemical reaction under compression. The parameter $V_{\#}(P)$ is directly linked to the pressure related apparent fragility (the steepness index, [3, 4, 28]) in the previtreous domain, namely:

$$m_T(P) = \frac{d\log_{10}\tau(T)}{d(P/P_g)} = \frac{P_g}{\ln 10}\frac{d\ln\tau(T)}{dP} \qquad (4)$$

As shown very recently, the apparent fragility exhibits the presumably 'universal' previtreous behavior on approaching the glass transition [28]:

$$m_T(P) = \frac{A}{P^* - P} \qquad (5)$$

where $T = const$, $P < P_g$, $P^* > P_g$ denotes the singular 'spinodal' pressure and $A = const$

To conclude the issue of $V_a(P)$ behavior, below the new and reliable way of its estimation is proposed. The subsequent application for experimental data confirms the that $V_{\#}(P) \neq V_a(P)$. To approach such conclusions, first the general SA-SB is considered [4]:

$$\tau(T,P) = \tau_0 \exp\left(\frac{PV_a(P) + E_a(T)}{RT}\right) \qquad (6)$$

For the isobaric, temperature related case one obtains:

$$\tau(T) = \tau_0 \exp\left(\frac{C}{RT}\right)\exp\left(\frac{E_a(T)}{RT}\right) = \tau_{ref.} \exp\left(\frac{E_a(T)}{RT}\right) \qquad (7)$$

where $P = const$, $C = V_a(P) \times P = const$ and the subscript 'ref.' denotes 'reference'.

For studies under atmospheric pressure one can approximate $P \sim 0$ and then $\tau_{ref.} \approx \tau_0$ what leads to the SA eq. (1). For studies under high pressures $\tau_{ref.} > \tau_0$ and the prefactor exhibits a weak temperature dependence. Regarding the isothermal and pressure-related behavior:

$$\tau(P) = \tau_{ref.}(T) \exp\left(\frac{V_a(P) \times P}{RT}\right), \qquad (8)$$

where $T = const$ and $\tau_{ref.} = \tau_0 \exp(E_a/RT)$, if the reference studies under atmospheric pressure are considered.

The last equation coincides with the basic SB eq. (2), and indicates that $10^{-10} < \tau_{ref.} < 10s$ for different tested isotherms [4, 5, 15-26]. In practice, $\tau_{ref.}$ can be easily determined from studies under atmospheric pressure. The fact that $\tau_{ref.}$ is well-known makes enable the following simple and reliable estimation of the apparent activation volume, directly from eq. (8):

$$V_a(P) = \frac{RT}{P} \ln\left(\frac{\tau(P)}{\tau_{ref.}(T)}\right), \qquad T = const \qquad (9)$$

Below experimental results testing eq. (9) are presented, basing on experimental measurements of the structural (primary, alpha) relaxation time $\tau(P)$ in few glass formers between the atmospheric pressure $\tau(P = 0.1 MPa)$ and the glass transition pressure $\tau(P_g) = 100 s$. Results are for glass forming low molecular weight diisobutyl phthalate ($T_g(0.1 MPa) = 196.8 K$), epoxy resin bisphenol A/epichlorohydrin (EPON 828, $T_g(0.1 MPa) = 253.9 K$) and liquid crystalline isoctyloxycyanobiphenyl (8*OCB, $T_g(0.1 MPa) = 220.7 K$. The latter vitrifies in the isotropic liquid phase and the possible nematic phase is hidden below the glass transition. These results were obtained using the broad band dielectric spectroscopy and structural relaxation times were determined from peak of primary relaxation loss curves ($\varepsilon''(f)$) from peak frequencies $\tau = 1/2\pi f_{peak}$. Experimental details are given in ref. [28].

Results of mentioned dielectric relaxation time measurements are presented in Figure. 1

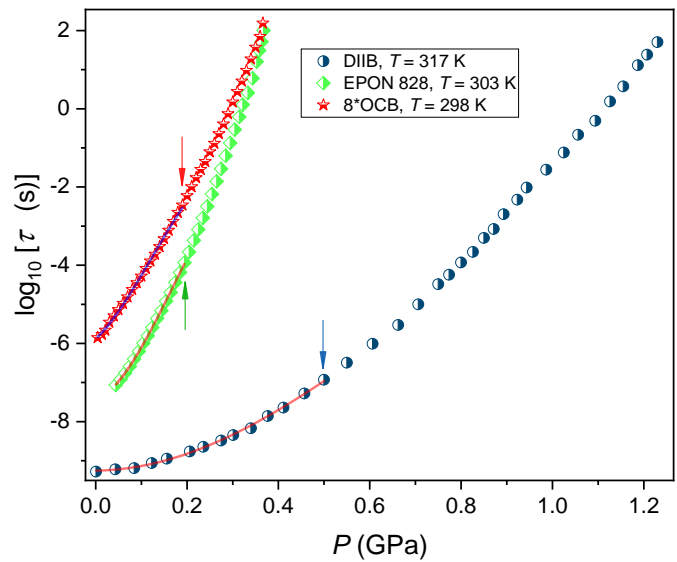

**Figure 1**     The isothermal pressure evolution of the structural relaxation time in previtreous domains between the (almost) atmospheric pressure ($P = 0.1 MPa$) and the glass transition $\tau(T_g, P_g) = 100 s$ for glass forming diisobutyl phthalate (DIIP), isoctyloxycyanobiphenyl (8*OCB), bisphenol A/epichlorohydrin (EPON 828). Solid curves show results of fittings via eq. (10), in the low/moderate range of pressures.

Figures (2), (3) and (4) present evolutions of calculated $V_a(P)$ apparent activation volumes (eq. (9)) and related to the apparent fragility $V^\#(P) \sim m_T(P)$ (eq. 4). Insets show reciprocals of the latter following the 'universal' pattern indicated via eq. (5), discussed in details in ref. [28].

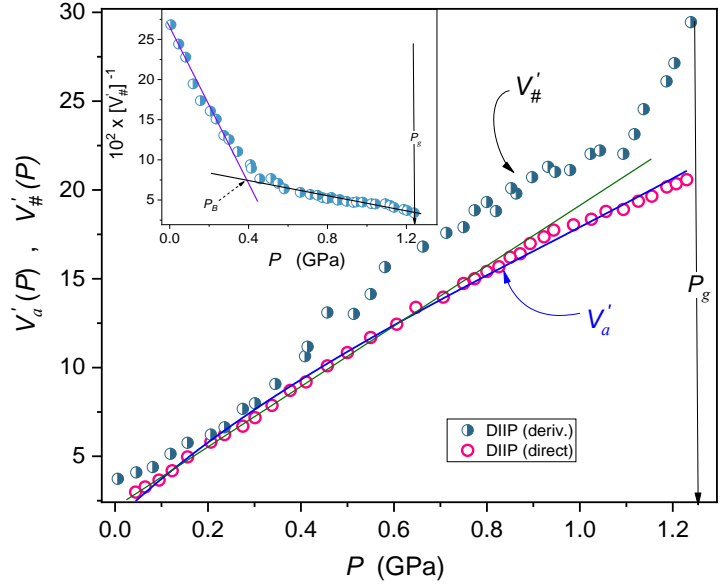

**Figure 2** The pressure evolution of the normalized apparent activation volume $V'_a(P)=V_a(P)/RT$ (eq. (9)) and $V'_\#(P)=V_\#(P)/RT \propto m_T(P)$ (eq. (4)). The inset shows the reciprocal of the latter (see eq. (5)). The blue curve is related for eq. (13), with parameters given in Table I. The green line shows the low/moderate pressures behavior. The dashed arrow in the inset indicates dynamic crossover relaxation time related to the 'universal' time scale $\tau(T_B)=10^{-7\pm1}s$ [28, 29]. Results are for diisobutyl phtalate, $T=317$ K.

It is clearly visible that $V^\#(P) \sim V_a(P)$ only for $P \to 0$. On approaching the glass transition $V^\#(P) \gg V_a(P)$. Considering the possible parameterization of the apparent activation volume one can note that ca. up to 200 MPa for 8*OCB and EPON 828 and even above 400 MPa for DIIP one can approximate: $V_a(P)=a+bT$. Substituting, the latter dependence into the SB eq. (8) one obtains:

$$\tau(P)=\tau_{ref.}\, exp\left(\frac{aP+bP^2}{RT}\right), \qquad T=const \tag{10}$$

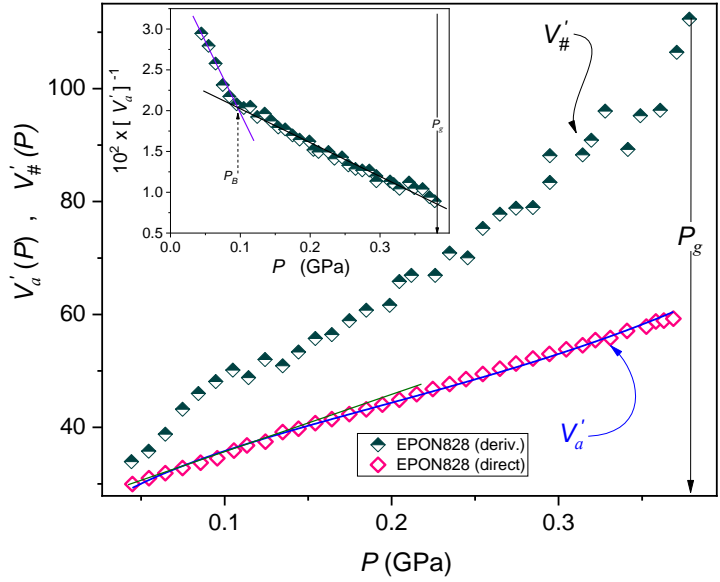

**Figure 3** The pressure evolution of the normalized apparent activation volume $V'_a(P) = V_a(P)/RT$ (eq. (9)) and $V'_\#(P) = V_\#(P)/RT \propto m_T(P)$ (eq. (4)). The inset shows the reciprocal of the latter (see eq. (5)). The blue curve is related for eq. (13), with parameters given in Table I. The green line shows the low/moderate pressures behavior. The dashed arrow in the inset indicates dynamic crossover relaxation time related to the 'universal' time scale $\tau(T_B) = 10^{-7\pm1}\,s$ [28, 29]. Results are for the previtreous behavior in liquid crystalline 8*OCB, $T = 298\,\text{K}$.

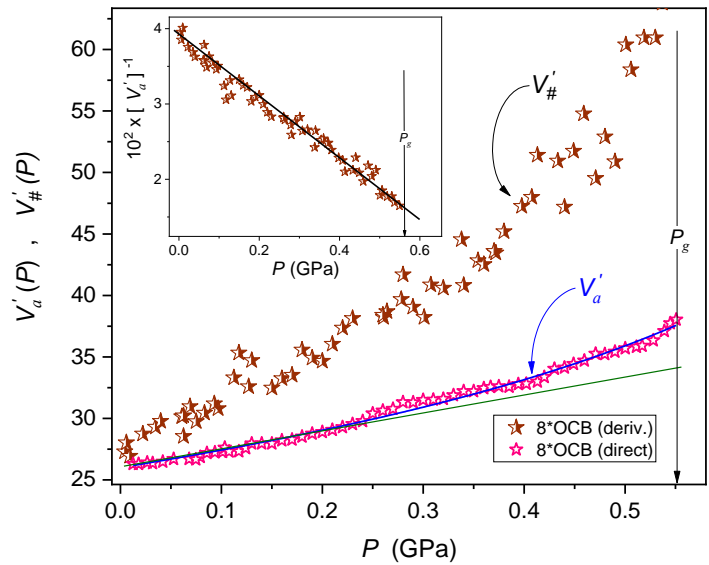

**Figure 4** The pressure evolution of the normalized apparent activation volume $V'_a(P) = V_a(P)/RT$ (eq. (9)) and $V'_\#(P) = V_\#(P)/RT \propto m_T(P)$ (eq. (4)). The inset shows the reciprocal of the latter (see eq. (5)). The blue curve is related for eq. (13), with parameters given in Table I. The green line shows the low/moderate pressures behavior. Results are for the previtreous behavior in EPON 828, $T = 303$ K.

Solid curves in Fig. 1 confirms the validity of eq. (10) for portraying $\tau(P)$ behavior in low/moderate pressure domains. Recently, Kornilov et al. [30] indicated that experimental reaction rates constants for pressures below 200-300 MPa can be reliably described by the relation $ln\, k(P) = a'P + b'P^2$, in fair agreement with the approximation given in eq. (10). When considering the possibility of portraying $V_a(P)$ behavior in the whole tested range of pressures one can recall the relation derived directly from eqs. (4), (5) [28]:

$$\tau(P) = \tau_{0P}(P^* - P)^{-\Phi} \tag{11}$$

Comparing eq. (11) and the SB eq. (2) or (8) one obtains the following relation:

$$V'_a(P) = \frac{V_a(P)}{RT} = \frac{C'}{P} - \frac{\Phi}{P} ln(P^* - P) \tag{12}$$

where $C' = ln\,\tau_{0P}/ln\,\tau_{ref}$.

Unfortunately, eq. (12) exhibits notable fitting problems for lower pressures, where $P \to 0$. However, for solids and liquids the 'terminal' pressure $P = 0$ can be smoothly 'cross-overed' into the negative pressures (isotropically stretched) domain, without any hallmarks when shifting $P > 0 \to P < 0$ [31]. Solids and liquids can be isotropically stretched down the absolute stability limit spinodal $P_S$, where the metastable system 'breaks' and returns to the equilibrium state under atmospheric pressure [31]. All these indicates that for solids and liquids the optimal pressure reference constitutes the stability limit under negative pressures. This suggest the following transformation: $P \to \Delta P = P - P_S$. Including this into eq. (12) one obtains:

$$V_a'(P) = \frac{C'}{\Delta P} - \frac{\Phi}{\Delta P} \ln(P^* - P) \tag{13}$$

Assuming values for singular pressures from the analysis of $1/m_T(P)$ [28] or $1/V_\#(P)$ (insets in Figs. 2, 3, 4, see also eq. (4)) one obtains a superior portrayal of $V_a'(P)$ experimental data, as shown by solid lines in Figures 2, 3, and 4. All related parameters are given in Table I.

**Table I.** Results of fitting of $V_a' = V_a/RT$ via eq. (13) for experimental data from Figures 2, 3 and 4 where they are shown as solid, blue, curves.

| Glass former | $P^*$ (GPa) | $C'$ (GPa$^{-1}$) | $\Phi$ | $P_S$ (GPa) |
|---|---|---|---|---|
| DIIB | 2.6 | 72.3 | 74 | -1.1 |
| EPON 828 | 0.6 | -12.1 | 27 | -0.1 |
| 8*OCB | 1.0 | 5.5 | 28.9 | -0.21 |

Notable is the determination of the absolute stability limit ($P_S$) what always constitute a difficult experimental task. Values of the exponent $\Phi$ reported in Table I agrees with ones obtained when applying eq. (11) for describing $\tau(P)$ experimental data in ref. [28].

Concluding, this reports proposes the new solution for the long standing problems of the apparent activation volume in complex systems. It indicates the simple and reliable way of estimating the apparent activation volume $V_a'(P)$ in the previtreous domain and the possibility of its parameterization. The latter includes the concept of negative pressures domain and offers the estimation of the absolute stability limit spinodal pressure. Results presented focus on the behavior of the primary relaxation time, but they can be applied also for the viscosity, electric conductivity, diffusion, equilibrium, reaction rates coefficients, ... what indicates the broad range of fundamental and practical applications ranging from the glass transition physics and the solid state physics to 'extreme' chemistry, geophysics and material engineering.

**Acknowledgement**

This research was carried out due to the support of the National Centre for Science (Poland), project NCN OPUS ref. 2016/21/B/ST3/02203, head Aleksandra Drozd-Rzoska.